\documentclass[12pt]{article}

\usepackage[margin=1in]{geometry}
\usepackage{amsmath,amssymb,graphicx,color}
\usepackage[titletoc,title]{appendix}
\usepackage{bm}
\usepackage{mathrsfs}
\usepackage{authblk}

\newcommand{\hzero}{h^{(0)}}
\newcommand{\hone}{h^{(1)}}
\newcommand{\htwo}{h^{(2)}}
\newcommand{\hthree}{h^{(3)}}

\newcommand{\xzero}{x^{(0)}}
\newcommand{\xone}{x^{(1)}}
\newcommand{\xtwo}{x^{(2)}}
\newcommand{\xthree}{x^{(3)}}

\newcommand{\Hkern}{\mathcal{H}}

\newcommand{\sdone}[2]{\frac{\mathrm{d} #1}{\mathrm{d} #2}}
\newcommand{\sdd}[3]{\frac{\mathrm{d}^{#1} #2}{\mathrm{d} #3^{#1}}}
\newcommand{\pd}[2]{\frac{\partial #1}{\partial #2}}
\newcommand{\pdd}[3]{\frac{\partial^{#1} #2}{\partial #3^{#1}}}

\newcommand{\Me}{M\kern-.15em e \,}
\newcommand{\cc}{\mathrm{c}.\mathrm{c}.}
\newcommand{\mye}{\text{e}}
\newcommand{\myi}{\text{i}}

\newcommand{\Nbar}{\overline{N}}

\usepackage{amsthm}

\title{Collective vibrations of a hydrodynamic active lattice}
\author{Stuart J. Thomson, Matthew Durey, Rodolfo R. Rosales}
\affil{Department of Mathematics, Massachusetts Institute of Technology, Cambridge, MA, 02139}
\date{}
\begin{document}
\maketitle

\begin{abstract}
Recent experiments show that quasi-one-dimensional lattices of self-propelled droplets exhibit collective instabilities in the form of out-of-phase oscillations and solitary-like waves. 
This hydrodynamic lattice is driven by the external forcing of a vertically vibrating fluid bath, which invokes a field of subcritical Faraday waves on the bath surface, mediating the spatio-temporal droplet coupling.
By modelling the droplet lattice as a memory-endowed system with spatially nonlocal coupling, we herein rationalise the form and onset of instability in this new class of dynamical oscillator. We identify the memory-driven instability of the lattice as a function of the number of droplets, and determine equispaced lattice configurations precluded by geometrical constraints. Each memory-driven instability is then classified as either a super- or sub-critical Hopf bifurcation \emph{via} a systematic weakly nonlinear analysis, rationalising experimental observations. We further discover a previously unreported symmetry-breaking instability, manifest as an oscillatory-rotary motion of the lattice. Numerical simulations support our findings and prompt further investigations of this nonlinear dynamical system.
\end{abstract}

\section{Introduction}
Classifying emergent properties of many-body systems, both passive and active, is a central theme of contemporary soft matter physics \cite{ramaswamy2010mechanics,marchetti2013hydrodynamics}. In recent years, myriad systems have emerged whose complex dynamics at the macroscale originate from the properties of their constituent particles, such as their shape, polarity, and activity or locomotion. Examples include the complex flows arising from active stresses exerted by bacteria suspended in fluid \cite{dunkel2013fluid,saintillan2013active}; vortex generation and nonlinear wave propagation in colloidal fluids \cite{bricard2013emergence,bricard2015emergent,souslov2017topo,geyer2018sounds}; and phonon-like excitations in crystals of particles and droplets confined to microfluidic channels \cite{beatus2006phonons, baron2008hydrodynamic, janssen2012collective, schiller2015collective, tsang2018activity}. In non-fluidic systems, theoretical models of active nonlinear lattices are shown to exhibit instabilities in the form of out-of-phase oscillations and solitary waves \cite{ebeling2000nonlinear, makarov2000soliton, dunkel2002coherent, chetverikov2006dissipative,nekorkin2012synergetic, chetverikov2018dissipative}, prompting experimental analogues in the form of active electronic circuits \cite{hirota1973theoretical, singer1999circuit, makarov2001dissipative,kotwal2019active}.

\begin{figure}
\includegraphics[width=\textwidth]{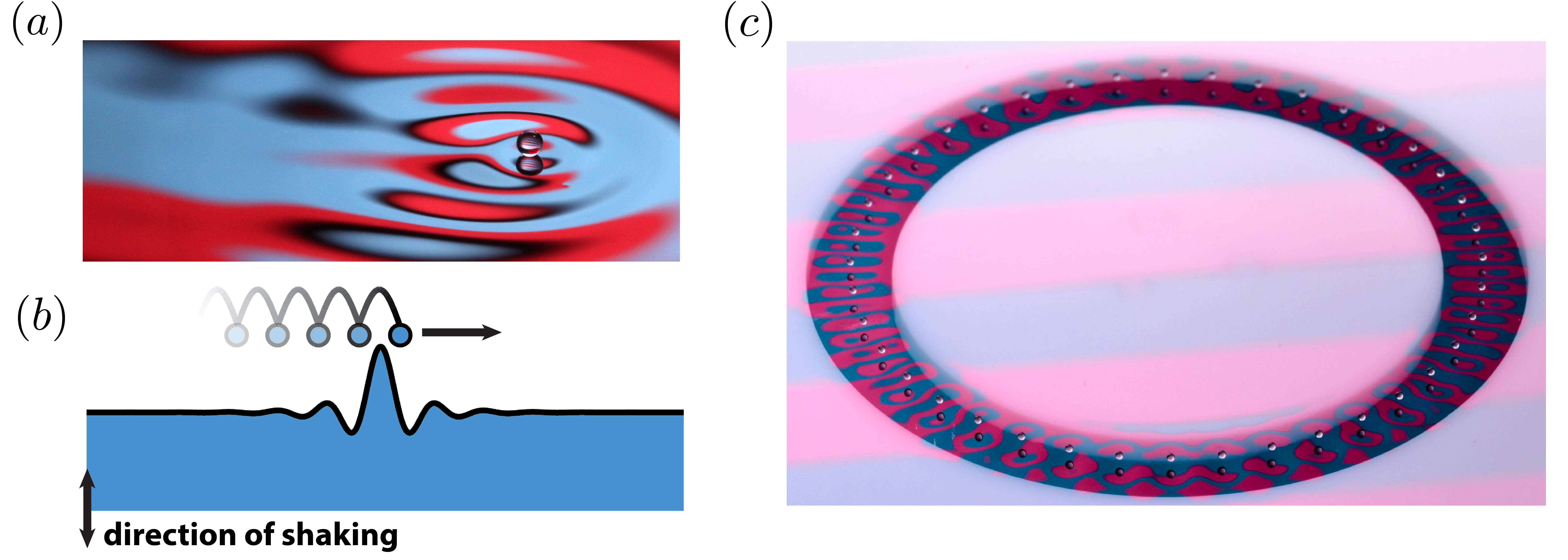}
\caption{(a) Single droplet of silicone oil self-propelling on the surface of a vibrating fluid bath. Picture courtesy of Daniel Harris. (b) Schematic of the droplet self-propulsion. Each time the droplet impacts the bath, it experiences a lateral force from the slope of the waves excited on the surface at previous impacts. (c) Oblique perspective of a chain of 40 equispaced droplets of silicone oil confined to a submerged annular channel, surrounded by a shallow layer of fluid.}
\label{fig:exp}
\end{figure}

We here focus on rationalising the instability of a new class of active oscillator \cite{thomson2020collective}, whose active units are self-propelled millimetric droplets bouncing on the surface of a vertically vibrating bath of viscous fluid \cite{couder2005dynamical,bush2015pilot}. In the absence of drops, the fluid interface remains flat below the Faraday threshold, the critical vibrational acceleration at which Faraday waves spontaneously appear on the bath surface. When placed on the surface, a millimetric droplet may bounce periodically in resonance with its self-generated, subcritical, subharmonic Faraday wave field (with a characteristic wavelength $\lambda_{F}$), exciting waves at each impact. Above a critical vibrational acceleration, the droplet destabilises to small, lateral perturbations. In this regime, dissipative effects due to drag are overcome by propulsive forces enacted by the slope of the droplet's guiding wave field, its so-called pilot wave (Figures \ref{fig:exp}(a) and (b)). Herein lies the active component of the droplet motion: self-propulsion is achieved by continual exchange of energy of the droplet with its environment, in this case the vibrating bath. Moreover, as the bath's vibrational acceleration is increased progressively, the decay time of the pilot wave lengthens and the droplet is thus influenced by more of its past, increasing the so-called memory time of the droplet's motion \cite{protiere2006particle}. Memory and self-propulsion are thus intimately connected, the former regulating the strength of the propulsive wave force required to overcome dissipation.

As demonstrated in \cite{thomson2020collective}, when confined to a submerged annular channel, droplets may couple to form an effectively one-dimensional lattice (Figure \ref{fig:exp}(c)). Each droplet in the lattice bounces in periodic synchrony with period $T_F$, generating a quasi-monochromatic wave field whose superposition over all the droplets mediates the spatio-temporal coupling of the lattice. For sufficiently large forcing, experiments show that the droplet lattice exhibits collective vibrations in the form of small-amplitude out-of-phase oscillations or solitary-like waves, depending on the proximity of neighbouring droplets. Moreover, the transition to each of the aforementioned states may be characterised by the onset of a super- and sub-critical Hopf bifurcation, respectively. 

Oscillations of the lattice emerge from the competition between droplet self-propulsion---arising through the interaction of each droplet with the slope of its own wave field---and wave-mediated coupling between droplets. This latter phenomenon represents a distinguishing feature of this new class of coupled oscillator: the waves produced at each droplet impact with the bath give rise to an effective self-generated, dynamic coupling potential between droplets (\emph{vis-\`a-vis} lattices subject to an externally-applied potential). The existence of this dynamic potential has far-reaching consequences; for example (i) there exist lattice equilibria in which the droplets sit at the \emph{local maxima} of the potential and (ii) the dynamic potential encodes the memory of the system, storing information regarding each droplets' past trajectory. While oscillators containing spatial and temporal nonlocality have received significant theoretical attention \cite{kim1997multistability,kuramoto2002coexistence, abrams2004chimera, abrams2006chimera,sethia2008clustered}, experimental, mechanical analogues are relatively scarce, limited to networks of coupled pendula \cite{martens2013chimera,wojewoda2016smallest}. 

Through a systematic analysis of the simplest possible model of the hydrodynamic lattice---introduced in \S\ref{sec:model}---we herein in seek to delineate the conditions under which qualitatively different bifurcations can arise, elucidating the key mechanims underpinning the emergent, collective properties of this new class of dynamical oscillator. In \S\ref{sec:ls}, we identify the memory-driven instability of the lattice, as well as equispaced configurations rendered uniformly unstable at all memory by the form of the lattice wave field. We rationalise the latter in the low-memory regime through an energy-like quantity that the system attempts to minimise. A weakly nonlinear analysis (\S\ref{sec:wnl}) then prescribes to each memory-driven instability a super- or sub-critical Hopf bifurcation, rationalising experimental observations \cite{thomson2020collective}. Further, we show that symmetry-breaking leads to a previously unreported self-induced, oscillatory-rotary motion of the lattice. Our analysis concludes in \S\ref{sec:vary} with an exploration into how the transition to each of the aforementioned instabilities is controlled by the droplet separation distance. We note that, while our study is focused on the regime where droplet motion is confined to an annular channel, the results of \S\S\ref{sec:model}--\ref{sec:vary} are also appropriate for describing azimuthal oscillations of droplet rings in free-space \cite{couchman2020rings}, allowing us to rationalise behaviour in this related system. Finally, the results of our paper are summarised in \S\ref{sec:con}, along with proposals for future work.

\section{Active lattice model}
\label{sec:model}

\begin{figure}
\begin{center}
\includegraphics[width=0.9\textwidth]{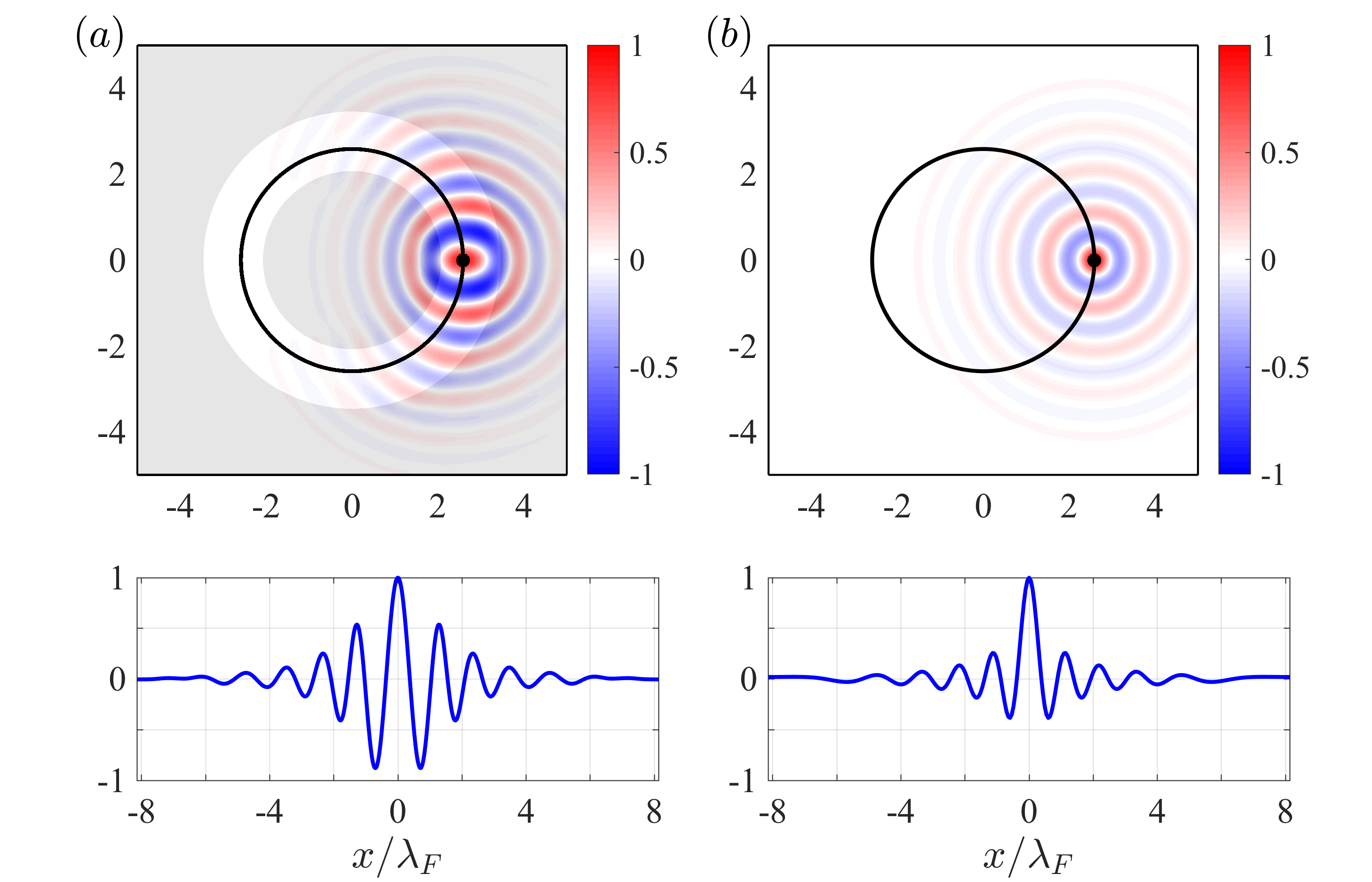}
\end{center}
\caption{$(a)$ An example wave field $\overline{\Hkern}(\bm{x};\bm{x}_p)$, corresponding to a stationary, bouncing droplet at $\bm{x} = \bm{x}_p$, computed using the model of Durey \emph{et al.}\ \cite{durey2020faraday} for $R/\lambda_F = 2.59$ (upper panel). The fluid and geometry parameters are listed in the Supplementary Material. Grey regions denote the shallow layer of fluid, and the deeper, annular channel has a white background. The position $\bm{x}_p$ is denoted by the black dot. Horizontal lengths are normalised by the Faraday wavelength, $\lambda_F$, and the vertical scale is normalised by $\overline{\Hkern}(\bm{x}_p; \bm{x}_p)$.
The radial cut $|\bm{x}| = R$ (denoted by the black circle in the upper panel) of $\overline{\Hkern}$ is presented in the lower panel, parametrised by the arc length, $x$, where $x = 0$ corresponds to the horizontal position $\bm{x} = \bm{x}_p$.
$(b)$ The analogous result to (a) for the wave field $\overline{\Hkern}(\bm{x};\bm{x}_p) =   F_0(|\bm{x}-\bm{x}_p|)$ defined by \eqref{eqn:wavefield}.
}
\label{fig:wavefield}
\end{figure}

We begin by constructing a model of the droplet lattice, valid below and near to the onset of instability, by first introducing some simplifying assumptions. In the physical regimes of interest, the time scale of the droplets' vertical motion, $T_{F}$, is typically much less than the time scale of horizontal motion. We thus approximate the wave field generated at impact by each droplet as the time-periodic wave field generated by a single, stationary, bouncing droplet at a horizontal position $\bm{x} = \bm{x}_p$. The wave field is normalised to account for the exponential decay time, $T_M$, of the Faraday waves \cite{couchman2019bouncing}. We next exploit the periodicity (and synchrony) of the droplets' vertical motion to average the droplets' horizontal motion over one bouncing period, $T_F$, giving rise to the stroboscopic wave field kernel $\overline{\Hkern}$ \cite{molavcek2013dropsb, oza2013trajectory}. We note that the presence of the annular subsurface topography in experiments means that $\overline{\Hkern}$ is not translationally invariant; the wave kernel depends on where the droplet resides in the channel, $\bm{x}_p$, and so $\overline{\Hkern} = \overline{\Hkern}(\bm{x}; \bm{x}_p)$. However, the geometry of the system is rotationally invariant; thus, in polar coordinates (whose origin coincides with the centre of the annulus) we have $\overline{\Hkern} = \overline{\Hkern}(r; r_{p}; \theta - \theta_{p})$ (see Figure \ref{fig:wavefield}(a)). The linear superposition of the waves generated by $N$ droplets yields the global wave field, the dynamic potential, which mediates the spatio-temporal coupling of the lattice.

The droplets are propelled by the wave field generated over all prior impacts and their motion is countered by a linear drag \cite{molavcek2013dropsb}. For $N$ droplets at positions $\bm{x}_{1}(t),\ldots,\bm{x}_{N}(t)$ and global wave field $\overline{h}(\bm{x},t)$, amalgamating the foregoing assumptions yields the many-droplet stroboscopic model \cite{oza2013trajectory,oza2017orbiting}, generalised to account for submerged topography:
\begin{subequations}
\label{eqn:2d_strobe}
\begin{align}
m\ddot{\bm{x}}_{n} + D\dot{\bm{x}}_{n} &= - mg \nabla \overline{h}(\bm{x},t)\vert_{\bm{x} = \bm{x}_{n}},\\
\pd{\overline{h}}{t} + \frac{1}{T_{M}}\overline{h} &= \frac{1}{T_{F}}\sum_{m = 1}^{N} \overline{\Hkern}(\bm{x}; \bm{x}_{m}),
\end{align}
\end{subequations}
where $m$ is the droplet mass, $D$ is a drag coefficient, $g$ is acceleration due to gravity, and dots denote differentiation with respect to time $t$.

In experiments, it is observed that the deviation from circumferential droplet motion is small \cite{thomson2020collective}. Using this fact, we recast the two-dimensional horizontal droplet motion in terms of the arc length $x$ along a circle of constant radius $R$ (which, in experiments, is controlled by the radius of the submerged channel). The projection of the stroboscopic bouncer wave field, $\overline{\Hkern}$, onto this circle is then $\mathcal{H}(x) = \overline{\Hkern}(R\bm{e}_r(x); R\bm{e}_r(0))$, where the radial unit vector $\bm{e}_r(x)$ is defined as $\bm{e}_r(x) = (\cos(x/R), \sin(x/R))$. Further, the rotational invariance of the system renders $\Hkern = \Hkern(x - x_{p})$, where the wave kernel $\mathcal{H}(x)$ is periodic with period $L = 2\pi R$. The model \eqref{eqn:2d_strobe} then becomes
\begin{subequations}
\label{eqn:strb_dim}
\begin{align}
\label{eqn:strb_4}
m \ddot{x}_{n} + D\dot{x}_{n} &= -mg \frac{\partial h}{\partial x}\Big\vert_{x = x_{n}}, \\
\frac{\partial h}{\partial t} + \frac{1}{T_M}h &= \frac{1}{T_F}\sum_{m = 1}^{N} \mathcal{H}(x - x_{m}). \label{eqn:strb_5}
\end{align}
\end{subequations}

Our model is closed by selecting a particular form of the wave kernel $\mathcal{H}(x) = \mathcal{H}(x+L)$. In principle, $\Hkern$ may be generated using the fluid mechanical model of Durey \emph{et al.}\ \cite{durey2020faraday}, in which the fluid evolution over the submerged annular channel is explicitly calculated (see Figure \ref{fig:wavefield}(a)). However, there are many subtle aspects of the wave field that obfuscate the key mechanisms underlying the dynamics of the hydrodynamic lattice. These include variations in the wave-field amplitude arising due to the particular bouncing phase of the drops \cite{couchman2019bouncing}, and memory-dependent changes to the exponential decay-length of the pilot wave \cite{milewski2015faraday, durey2017faraday, turton2018review,tadrist2018faraday}. Our theory is instead developed for a general, flexible, periodic wave-kernel, $\mathcal{H}(x)$. For exploration and demonstration purposes, we simply define a wave kernel that exhibits the fundamental aspects of the fluid system: namely, a quasi-monochromatic wave field endowed with exponential spatial-decay. 
Moreover, as arises from the physics of each bouncing droplet \cite{molavcek2013dropsb, oza2013trajectory}, the stroboscopic wave kernel, $\Hkern$, exhibits a peak at the droplet position, which ultimately leads to stable lattice configurations wherein each droplet sits at a local maximum of the dynamic potential, $h$.
An application---using the results of this study---to the specific fluid system explored in experiments is subject to future work.

To inform our choice of candidate wave kernel, $\mathcal{H}(x)$, we observe numerically that topographically induced deviations in $\overline{\Hkern}(\bm{x};\bm{x}_p)$ from an axisymmetric wave kernel  about $\bm{x}_p$ are weak (see Figure \ref{fig:wavefield}(a)). For simplicity, we thus consider $\overline{\Hkern}(\bm{x};\bm{x}_p) = F_0(|\bm{x} - \bm{x}_p|)$ for our simulations, where a candidate axisymmetric wave form $F_0(r)$ is
\begin{equation}
\label{eqn:wavefield}
 F_0(r) = \mathcal{A}_0J_{0}(k_F r)\text{sech}(r/l_d).
\end{equation}
Here, $J_{0}$ is the zeroth-order Bessel function, $\mathcal{A}_0$ is the wave amplitude, $k_F = 2\pi/\lambda_F$ is the Faraday wavenumber, and $l_d$ is a tunable spatial decay-length, which in turn determines the strength of the nonlocal droplet-coupling. 
The wave kernel, $\mathcal{H}(x)$, is then defined as the radial cut of $\overline{\Hkern}(\bm{x};\bm{x}_p)$ along a circle of constant radius $R$, specifically
\begin{equation}
\label{eqn:dim_Hkern}
\mathcal{H}(x) = F_0\left(2 R \sin\frac{x}{2R}\right),
\end{equation} 
as demonstrated by the black circle in Figure \ref{fig:wavefield}(b). A further advantage of \eqref{eqn:dim_Hkern} is that it allows us to easily explore the implications of varying the lattice radius, $R$, continuously (\S\ref{sec:vary}). For peace of mind, we tested this generic wave kernel against the numerically computed wave kernel \cite{durey2020faraday}, which demonstrated sufficient qualitative similarities to still capture the dynamical motion of the active lattice. Finally, we emphasise that the analysis presented herein is independent of the particular choice of $\Hkern$, provided that $\mathcal{H}$ is periodic and sufficiently smooth, with oscillations arising on a length scale similar to the Faraday wavelength.

To non-dimensionalise \eqref{eqn:strb_dim}--\eqref{eqn:dim_Hkern}, we set $\lambda_F$ as the typical horizontal length-scale, $t_0 = m/D$ as the typical time scale, and $h_0 = \lambda_F^2/gt_0^2$ as the typical free-surface elevation. By rescaling $\mathcal{H} \mapsto \mathcal{H}_0 \mathcal{H}$, where $\mathcal{H}_0 = h_0 T_F/t_0$, the dimensionless model then reads
\begin{subequations}
\label{eqn:strb_dimensionless}
\begin{align}
\label{eqn:strb_6}
\ddot{x}_{n} + \dot{x}_{n} &= -\frac{\partial h}{\partial x}(x_{n},t),\\
\label{eqn:strb_7}
\frac{\partial h}{\partial t} + \nu h &= \sum_{m = 1}^{N} \mathcal{H}(x - x_{m}),
\end{align}
\end{subequations}
where
\begin{equation}
\label{eqn:wavefield_nd}
\mathcal{H}(x) = F\left(2 r_0 \sin\frac{x}{2r_0}\right),\quad F(r) = \mathcal{A}J_{0}(2\pi r)\text{sech}(r/l).
\end{equation}
The dimensionless parameters are $l = l_d/\lambda_F$, $r_0 = R/\lambda_F$, $\nu = t_{0}/T_M > 0$, and $\mathcal{A} = \mathcal{A}_0/\mathcal{H}_0$, while the dimensionless circumference of the lattice is $\Lambda = L/\lambda_{F} = 2\pi r_0$. Unless stated otherwise, for the numerical results presented herein, we fix $l = 1.6$, $\mathcal{A} = 0.1$, and $r_{0} = 5.4$, based on typical experimental values. 

To summarise, waves are generated about the current position of each droplet and superpose to form the global wave field, $h(x,t)$. The spatio-temporal evolution of $h(x,t)$ is regulated by \eqref{eqn:strb_7} and depends on the trajectories of each droplet, imprinting a path-memory on the system. Each droplet is then driven by the local gradient of $h(x,t)$ through \eqref{eqn:strb_6}. The effects of memory in our dimensionless system \eqref{eqn:strb_dimensionless} are encoded through the dimensionless dissipation rate $\nu\sim 1/T_{M}$. While $\nu$ is convenient algebraically, we will interpret our results using a more natural measure of memory $M = 1/\nu$, where larger $M$ means that past dynamics play a more prominent role.  

\section{Memory-driven and geometric instability}
\label{sec:ls}
To determine the critical value of the memory $M = M_{c}$ at which the wave force promotes self-propulsion of the droplets, we analyse the linear stability of \eqref{eqn:strb_dimensionless} to small perturbations about a static, equispaced lattice configuration, coinciding with experiments \cite{thomson2020collective}. Initially, the droplets are positioned at $x_n = n\delta$, where $\delta = \Lambda/N$. Equation \eqref{eqn:strb_7} then yields the corresponding free-surface elevation 
\begin{equation}
\label{eqn:ls1}
h(x) = h_0(x) =  \frac{1}{\nu} \sum_{m = 1}^{N} \Hkern\left(x - m\delta\right). 
\end{equation}
By symmetry, the free-surface gradient beneath each droplet vanishes in this steady configuration. Furthermore, we note from \eqref{eqn:ls1} that one effect of increasing the memory, $M = 1/\nu$, is to increase the wave amplitude of the steady lattice. Upon consideration of a small perturbation to each droplet position, $x_n$, and the concomitant perturbation to the wave field, $h$, we set
$$x_{n} = n\delta + \eta\hat{x}_{n},\quad h = h_0+ \eta\hat{h}, $$ 
where $0 < \eta \ll  1$.
By substituting this form into \eqref{eqn:strb_dimensionless} and linearising, we obtain
\begin{subequations}
\label{eqn:L1}
\begin{align}
\label{eqn:ls2}
\ddot{x}_n + \dot{x}_n &= -\left[\pd{h}{x} + x_n\pdd{2}{h_0}{x}\right]\bigg\vert_{x = n\delta},\\
\label{eqn:ls3}
\frac{\partial h}{\partial t} + \nu h &= -\sum_{m = 1}^{N}x_{m}\Hkern'(x - m\delta),
\end{align}
\end{subequations}
where we have dropped the carets on the perturbed variables, and primes denote differentiation with respect to $x$.

The difficulty in analysing \eqref{eqn:L1} rests in the temporal nonlocality of the free-surface perturbation, $h(x,t)$. To circumnavigate this issue and project the dynamics entirely onto the droplet trajectories, $x_n(t)$, we use the form of \eqref{eqn:ls3} to define the auxiliary variables $\overline{x}_{n}(t)$ satisfying
\begin{equation}
\label{eqn:ls4}
\dot{\overline{x}}_{n} +\nu \overline{x}_{n} = x_{n} \quad \mathrm{for}\quad n = 1,\ldots,N,
\end{equation}
a procedure that we also apply throughout the weakly nonlinear analysis presented in \S\ref{sec:wnl}. 
It then follows from \eqref{eqn:ls3} that the evolution of the perturbed free-surface elevation, $h(x,t)$, may now be expressed in terms of the auxiliary variables, $\overline{x}_n(t)$. Specifically, we obtain the particular solution
\begin{equation}
\label{eqn:ls5}
h = -\sum_{m = 1}^{N} \Hkern'(x - m\delta) \overline{x}_{m}.
\end{equation}
The homogeneous solution to \eqref{eqn:ls3} decays exponentially in time, and thus plays no role in either the linear or weakly nonlinear stability (\S\ref{sec:wnl}) of the system. Having expressed $h$ in terms of the auxiliary variables $\overline{x}_n$, we recast \eqref{eqn:L1} as a reduced dynamical system for the variables $x_n$ and $\overline{x}_n$.

Substituting \eqref{eqn:ls5} into \eqref{eqn:ls2}, and using \eqref{eqn:ls1}, we find that $\mathcal{L}_{n}(\bm{x}) = 0$, where $\bm{x} = (x_{1},\cdots,x_{N})$. The linear operator $\mathcal{L}_{n}$ is defined as
\begin{equation}
\label{eqn:ls6}
\mathcal{L}_{n}(\bm{x}) = \ddot{x}_{n} + \dot{x}_{n} + 
\sum_{m = 1}^N
\bigg(\frac{x_n}{\nu} - \overline{x}_{m}\bigg)
\Hkern''(\delta(n-m)),
\end{equation}
which, together with equations \eqref{eqn:ls4}, constitute a linear system of $2N$ ordinary differential equations describing the evolution of $x_{n}$ from the steady state. Equations \eqref{eqn:ls4} and \eqref{eqn:ls6} may be solved using standard eigenmode methods, the periodicity of the steady lattice prompting the \emph{ansatz}
\begin{equation}\label{eqn:ls7}
x_{n} = A\exp\left(\myi kn\alpha + \lambda_{k}t\right) + \cc ,\quad \overline{x}_{n} = \frac{A}{\lambda_k + \nu} \exp\left(\myi kn\alpha + \lambda_{k}t\right) + \cc,
\end{equation}
where the form of $\overline{x}_{n}$ follows from \eqref{eqn:ls4}.  Here we have defined the angular spacing parameter $\alpha = 2\pi/N$, imaginary unit $\myi$, and complex amplitude $A$, where $\cc$ denotes complex conjugation of the preceding term. By symmetry considerations, we restrict our attention to the wavenumbers $k = 0,\ldots,\Nbar$, where $\Nbar = \lfloor N/2 \rfloor$. By substituting \eqref{eqn:ls7} into \eqref{eqn:ls6}, we obtain the dispersion relation $\mathcal{D}_k(\lambda_k; \nu) = 0$, where 
$$
\mathcal{D}_k(\lambda; \nu) = \lambda^2 + \lambda + \frac{c_0}{\nu} - \frac{c_k}{\lambda + \nu}.
$$
The real constants $c_k$ are defined as
\begin{equation}
\label{eqn:ck}
c_k = \sum_{n = 1}^{N}\cos\left(kn\alpha\right)\Hkern''(\delta n),
\end{equation}
which may be interpreted as the discrete cosine transform coefficients of the even and periodic function $\Hkern''(x)$, arising from the discrete convolution in equation \eqref{eqn:ls6}.

Upon rearranging $\mathcal{D}_k(\lambda_k; \nu) = 0$ and writing $\nu = 1/M$, the eigenvalues, $\lambda_k$, describing the asymptotic linear stability of \eqref{eqn:L1}, satisfy the cubic polynomial
\begin{equation}
\label{eqn:disp_rel}
M\lambda^3_{k} + (M+1)\lambda^2_{k} + (c_{0}M^2 + 1)\lambda_{k} + M(c_{0} - c_{k}) = 0,
\end{equation}
whose roots we typically compute numerically.
The uniform lattice is asymptotically unstable if, for some wavenumber $k$, there exists an eigenvalue $\lambda_k$ satisfying $\text{Re}(\lambda_k) > 0$. 
A fundamental property of this lattice is the rotational invariance, characterised by the eigenvalue $\lambda_0 = 0$, which may allow for a slow drift of the lattice in the weakly nonlinear regime (see \S\ref{sec:wnl}).

It transpires that the lattice may destabilise \emph{via} two distinct mechanisms: (i) a memory-independent instability, where the lattice wave field destabilises the uniform configuration for all memory $M > 0$ (so $M_{c} = 0$); and (ii) a memory-driven instability when the wave force exceeds the drag force, prompting propulsion of the droplets for $M > M_{c} > 0$ (to be determined). Physically, case (i) arises from geometrical frustration of the lattice wave field: for particular lattice configurations, the lattice wave field precludes the formation of a stable, equispaced lattice, forcing the droplets to occupy a nearby equilibrium configuration. Henceforth, we will thus refer to case (i) as a \emph{geometric instability}. The memory-driven instability (ii) may be further decomposed into two subcategories: instability \emph{via} a real eigenvalue, giving rise to the steady rotation of the lattice, analogous to the dynamics of a single walking droplet \cite{oza2013trajectory}; and oscillatory instabilities arising for droplets in close proximity, akin to those observed in experiments of this lattice system \cite{thomson2020collective}. We discuss each of the foregoing cases and their implications in the following three sections.

\subsection{Geometric instability}
\label{sec:c3}
For short memory $M\ll 1$ (equivalently, $\nu\gg1$), the polynomial \eqref{eqn:disp_rel} admits three real roots:
\begin{equation}
\label{eqn:nugg}
\lambda_{k}^{(1)} = -1 + O(M),\quad \lambda_{k}^{(2)} = -\frac{1}{M} + O(1),\quad\text{and}\quad \lambda_{k}^{(3)} = (c_{k} - c_{0})M + O(M^2).
\end{equation}
Thus, if there exists a wavenumber $ k_{*}$ such that $c_{k_{*}} > c_{0}$ then $\lambda_{k_*}^{(3)} > 0$ and the lattice is unconditionally unstable for $M\ll 1$. In fact, the lattice is unconditionally unstable for all $M > 0$ when $c_{k} > c_{0}$, the negativity of the constant term in \eqref{eqn:disp_rel} always giving rise to a positive real root. While the asymptotic results \eqref{eqn:nugg} provide a mathematical rationale for this memory-independent instability, they provide little in the way of physical intuition, which we presently provide.

\begin{figure}
\begin{center}
\includegraphics[width=\textwidth]{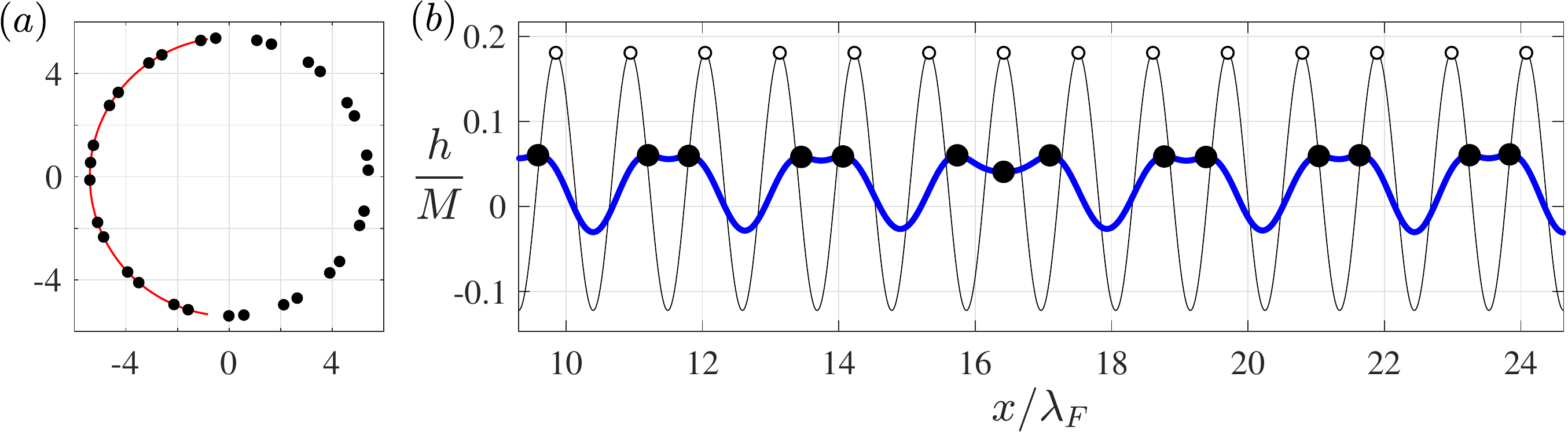}
\end{center}
\caption{The onset of geometric instability in the low-memory regime for $N = 31$ droplets. Black circles denote the final droplet rest state, a configuration that provides a local minimiser for the mean wave-height $\mathscr{H}$ (see \eqref{eqn:HScr}). (a) Overhead view of the final lattice, where lengths are normalised by the Faraday wavelength. (b) Initial (thin curve) and final (thick curve) lattice wave fields, $h$, plotted over the arc denoted by the red curve in (a). White circles denote the initial positions of the equispaced, unstable lattice.}
\label{fig:geo_inst}
\end{figure}

We first introduce the rescaled variables $\tau = Mt$ and $H = h/M$ and substitute into equations \eqref{eqn:strb_dimensionless}. We then consider the low-memory limit $M \ll 1$, leading us to neglect the highest-order time derivative in each equation. The result is overdamped, gradient-driven motion for each droplet described by

\begin{equation}
\label{eqn:lowmem}
\sdd{}{x_n}{\tau} = - \sum_{m=1}^{N} \mathcal{H}'(x_n - x_{m}),
\end{equation}

where the wave field is $H(x,\tau) = \sum_{m=1}^{N} \mathcal{H}(x - x_{m}(\tau))$. If all but one of the droplets were static, then the remaining droplet would evolve in response to a fixed potential prescribed by the wave field. However, due to the spatially nonlocal coupling of the lattice, this motion in turn alters the wave force acting on all other droplets, prompting propulsion. As such, it is necessary to consider the global dynamics of the system. 

We characterise the dynamics by the mean height of the wave field beneath each droplet
\begin{subequations}
\label{eqn:HScr}
\begin{equation}
\label{eq:Hscr_def_1}
\mathscr{H}(\tau) = \frac{1}{N}\sum_{n = 1}^N H(x_n(\tau), \tau) = \frac{1}{N}\sum_{n = 1}^N \sum_{m = 1}^N \Hkern(x_n(\tau) - x_m(\tau)).
\end{equation}

Using \eqref{eqn:lowmem} and that $\Hkern(x)$ is an even, periodic function, it is readily verified that

\begin{equation}
\label{eq:Hscr_def_2}
\sdd{}{\mathscr{H}}{\tau} = -\frac{2}{N}\sum_{n = 1}^N \bigg(\sdd{}{x_n}{\tau}\bigg)^2 \leq 0,
\end{equation}
\end{subequations}
indicating that the mean wave-height, $\mathscr{H}$, decreases along droplet trajectories (except at lattice equilibria). This phenomenon was also observed numerically in simulations of two-dimensional lattices \cite{couchman2020rings}. Indeed, the foregoing argument readily generalises to this latter case, and thus may be applied to this wider class of system \cite{protiere2005self,lieber2007self,eddi2009archimedean,eddi2011oscillating}.

We now provide an alternative interpretation of the geometric instability in terms of the energy-like quantity $\mathscr{H}$. When perturbing about the equispaced lattice, with $x_n(\tau) = n\delta + \eta \hat x_n(\tau)$ (with $0 < \eta \ll 1$), we compute the change in $\mathscr{H}$ from the steady state $\mathscr{H}_0 = \sum_{n = 1}^N\Hkern(\delta n)$. By substituting this \emph{ansatz} into \eqref{eq:Hscr_def_1} and Taylor expanding in powers of $\eta$, we obtain 
$$
\mathscr{H} = \mathscr{H}_0 + \eta^2\bigg[ \frac{1}{N} \hat{\bm{x}}^T (c_0 \bm{I} - \bm{A})\hat{\bm{x}}\bigg] + O(\eta^3), $$
where $\hat{\bm{x}} = (\hat x_1, \ldots, \hat x_n)^T$, $\bm{I}$ is the identity matrix, and $\bm{A}$ is the symmetric, circulant matrix with entries $\bm{A}_{nm} = \Hkern''(\delta(n-m))$. The stability of the equispaced lattice thus depends entirely on the definiteness of the matrix $(c_0 \bm{I} - \bm{A})$, where negative eigenvalues indicate instability since $\mathscr{H}$ can decrease in the direction of the corresponding eigenvector.  Due to the form of $\bm{A}$, its eigenvalues are $c_k$ (for $k =0, \ldots, N-1$) with corresponding eigenvectors $\bm{v}_k$ whose $n^\mathrm{th}$ element is $\mye^{\myi k n \alpha}$ (consistent with the normal modes \emph{ansatz} in \eqref{eqn:ls7}). Hence, the energy may decrease if there exists $k_*$ such that $c_0 - c_{k_*} < 0$ (recall \eqref{eqn:nugg}), rendering the steady lattice a saddle. When multiple wavenumbers satisfy this condition, the lattice rearranges in the direction of steepest descent of this energy-like quantity, $\mathscr{H}$, prescribed by the wavenumber $k_*$ that minimises $c_0 - c_{k}$.  When $N$ is even and $k_* = N/2$, symmetry dictates that the droplets approach a lattice whose separation distances alternate in size. For other values of $k_*$, more exotic, asymmetric lattice structures emerge, such as the 31-droplet lattice presented in Figure \ref{fig:geo_inst}. In fact, such an instability can even affect two well-separated droplets (specifically, $\delta\gg l$), despite the waves felt by the other droplet being exponentially small, emphasising the extent of the spatially nonlocal coupling.

\subsection{\label{case:c1} Onset of collective walking}

If the lattice configuration is not geometrically unstable, the system may destabilise to collective walking when the memory, $M$, increases above a critical threshold (to be determined). One such instability arises when a real eigenvalue of \eqref{eqn:disp_rel} passes through the origin, going from negative to positive.
Such an instability typically arises when $k = 0$ and $\lambda_0$ becomes a double eigenvalue at the instability threshold, although it is also possible in the exceptional case when there exists $k = k_{*}$ such that $c_{k_{*}} = c_{0}$ (since $\lambda_{k_*} = 0$ is then also a trivial root of \eqref{eqn:disp_rel}).
By factorising \eqref{eqn:disp_rel} when $c_{k} = c_{0}$, the non-trivial eigenvalues satisfy the quadratic polynomial
\begin{equation}
\label{eqn:ns_disp}
M \lambda^2_0 + (M + 1)\lambda_0 + \left(c_{0}M^2 + 1\right) = 0.
\end{equation}
Since $M > 0$, the stability of the system depends on the sign of the constant term in \eqref{eqn:ns_disp}, where $c_0$ determines the local curvature beneath each droplet in the steady lattice.

When $c_{0} < 0$, corresponding to each droplet sitting on a peak of the free surface, both roots of \eqref{eqn:ns_disp} are real. The system destabilises at the critical threshold $M =1/\sqrt{-c_{0}}$, at which one eigenvalue transitions from a stable node to a saddle, characteristic of a pitchfork bifurcation. Physically, this is the threshold at which the wave force dominates the drag force, promoting unidirectional self-propulsion, and thus is the many-droplet analogue of the walking threshold described in \cite{oza2013trajectory}, for which a supercritical pitchfork bifurcation arises \cite{protiere2006particle}. 

In contrast, when $c_{0} > 0$, each droplet sits in a trough of the free surface, a highly stable configuration. As such, $\text{Re}(\lambda_0) < 0$ for all $M$, where the system may return to its rest state \emph{via} either overdamped or underdamped oscillations. 
Such a regime may arise when the troughs about the central peak of the wave kernel, $\mathcal{H}$, are sufficiently deep (akin to the numerical example presented in Figure \ref{fig:wavefield}(a)) and the droplets lie in close proximity.

\subsection{Oscillatory instability}
\label{sec:hopf}
Of particular relevance to the experiments described in \cite{thomson2020collective} is the case where the lattice undergoes instead an oscillatory instability, wherein the real part of a complex-conjugate pair of eigenvalues transitions from negative to positive as memory is increased (indicative, but not conclusive, of a Hopf bifurcation; see \S\ref{sec:wnl}).
We note that a lattice may have consecutive pitchfork and oscillatory instabilities as $M$ is varied, in which case the form and threshold of the instability is determined by the smallest such critical $M$.
\begin{figure}
\begin{center}
\includegraphics[width=\textwidth]{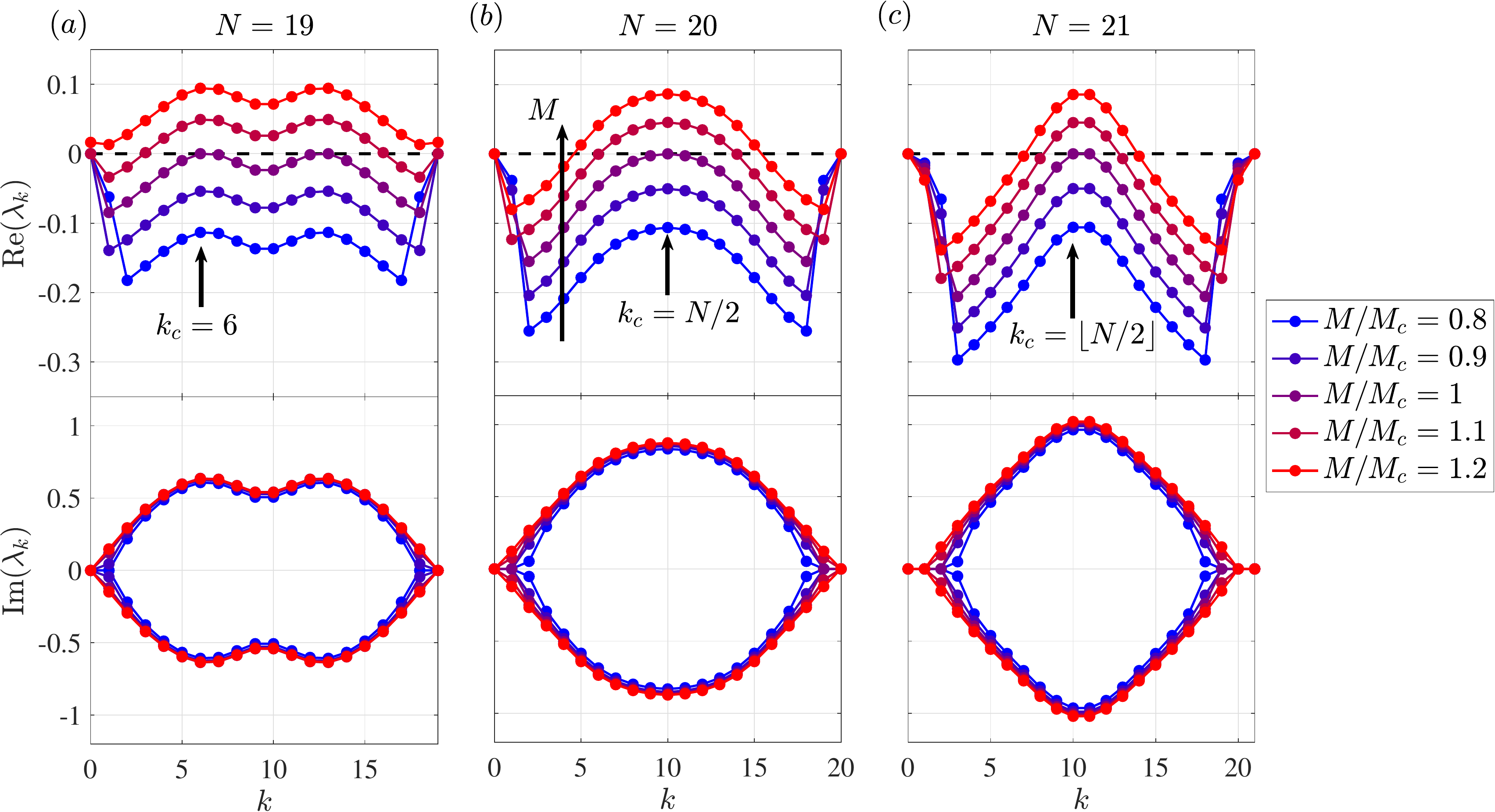}
\caption{The real (top panel) and imaginary (bottom panel) parts of the dominant eigenvalues, $\lambda_{k}$, determined from the cubic polynomial \eqref{eqn:disp_rel} for (a) $N = 19$, (b) $N = 20$, and (c) $N = 21$. As $M$ is gradually increased, a single wavenumber $k_{c}\in [0,\overline{N}]$ touches $\text{Re}(\lambda_{k_{c}}) = 0$, while $\text{Im}(\lambda_{k_{c}})\neq 0$. The upper and lower branches correspond to one member of the conjugate pair. As $M$ is increased further, yet more neighbouring wavenumbers destabilise about $k_{c}$. In contrast to $N = 20$ and $N = 21$, there is a notable departure of $k_{c}$ from $k = \overline{N}$ when $N = 19$, a hallmark of subcritical bifurcations~(see~\S\ref{sec:vary}).}
\label{fig:fig1}
\end{center}
\end{figure}

In Figure \ref{fig:fig1}, we demonstrate the possible forms of an oscillatory instability through consideration of the real and imaginary parts of the dominant eigenvalue, $\lambda_k$, for three consecutive droplet configurations: $N = 19$, $N = 20$, and $N = 21$.
As the memory, $M$, is increased, a single wavenumber $k_{c}\in[0,\overline{N}]$  touches $\text{Re}(\lambda_{k_{c}}) = 0$ for some critical value of $M = M_{c}$ (determined below), while $\text{Im}(\lambda_{k_{c}})\neq 0$, corresponding to an oscillatory instability. As $M$ is increased further, beyond $M_{c}$, yet more wavenumbers destabilise as their corresponding eigenvalues cross the imaginary axis. Of particular interest is the case $N = 20$ (Figure \ref{fig:fig1}(b)), where the critical wavenumber is $k_{c} = N/2$ in the current parameter regime. According to the linear stability analysis, each droplet position is described by $x_{n} = n\delta + (-1)^{n}[A\exp(\myi\omega_{k_{c}}t) + \cc]$ at $M = M_{c}$ in this case, corresponding to out-of-phase oscillations of adjacent droplets, akin to experimental observations \cite[Fig.\ 2]{thomson2020collective}. We also observe a notable departure in $k_{c}$ from $k = \overline{N}$ for $N = 19$ (Figure \ref{fig:fig1}(a)), contrasting with $N = 20$ and $N = 21$. This shift appears to be a hallmark of subcritical bifurcations (see \S\ref{sec:vary}). To fully describe the lattice dynamics for $M > M_{c}$, we consider the weakly nonlinear stability of these oscillations in \S\ref{sec:wnl}.

At the critical threshold $M = M_{k}$, the memory at which wavenumber $k$ goes unstable, the critical eigenvalue is $\lambda_{k} = \text{i}\omega_{k}$, where $\omega_k > 0$ without loss of generality. Upon substituting into \eqref{eqn:disp_rel} and taking real and imaginary parts, we obtain two equations for $\omega_{k}$, namely
\begin{equation}
\label{eqn:omega_find}
\omega^2_{k} = c_{0}M_{k} + \frac{1}{M_{k}}
\quad\mathrm{and}\quad 
\omega_k^2 = \frac{M_{k}(c_{0} - c_{k})}{M_k + 1}.
\end{equation}
By eliminating $\omega_k$, we find that $M_k$ satisfies $p(M_k) = 0$, where $p(M) = c_{0}M^3 + c_k M^2 + M + 1$. We identify $M_{k}$ as the smallest real root of $p$ for each wavenumber $k$ and define the critical lattice instability threshold as $M_c = \min_k M_k$, the minimum memory over all wavenumbers. The corresponding critical angular frequency $\omega_{c} = \omega_{k_{c}}$ may then be determined from \eqref{eqn:omega_find}, namely $\omega_c = \sqrt{c_{0}M_{c} + 1/M_{c}}$. 

\begin{figure}
\begin{center}
\includegraphics[width=\textwidth]{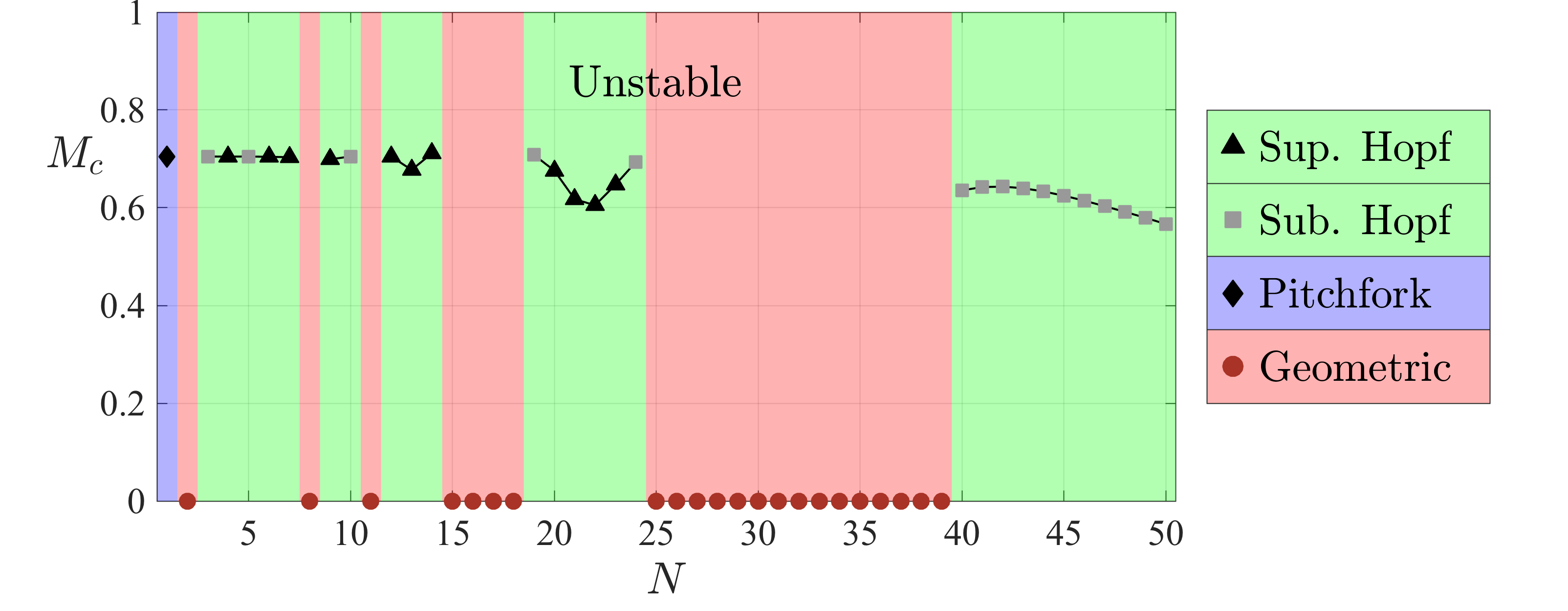}
\end{center}
\caption{The critical instability threshold, $M_c$, as a function of the number of droplets, $N$, for a fixed ring radius $r_{0} = 5.4$. The lattice is unstable when the memory parameter, $M = \nu^{-1}$, satisfies $M > M_c$, and is stable for $M < M_c$. The form of each instability is characterised by the legend. Black lines between data points are visual guides.
}
\label{fig:supsub}
\end{figure}

\subsection{Summary}
\label{sec:bif_summ}
The form of each instability outlined in \S\S\ref{sec:ls}\ref{sec:c3}--\ref{sec:hopf} is summarised in Figure \ref{fig:supsub}, where we present $M_{c}$ for each droplet configuration consisting of $N$ droplets. Lattice configurations are either geometrically unstable $(M_{c} = 0)$, or instability is triggered by memory effects ($M_{c} > 0$) through a pitchfork bifurcation or an oscillatory instability. (Foreshadowing the results of \S\ref{sec:wnl}, the oscillatory instabilities may be categorised as either supercritical or subcritical Hopf bifurcations). We observe no apparent pattern in the distribution of memory-driven or geometric instabilities, principally due to the fact that $N$ is varied discretely. As portrayed in \S\ref{sec:vary}, a more illuminating view is afforded when changing the lattice radius, $r_0$, continuously and fixing $N$.

\section{Weakly nonlinear oscillations and self-induced drift}
\label{sec:wnl}
\begin{figure}
\begin{center}
\includegraphics[width=\textwidth]{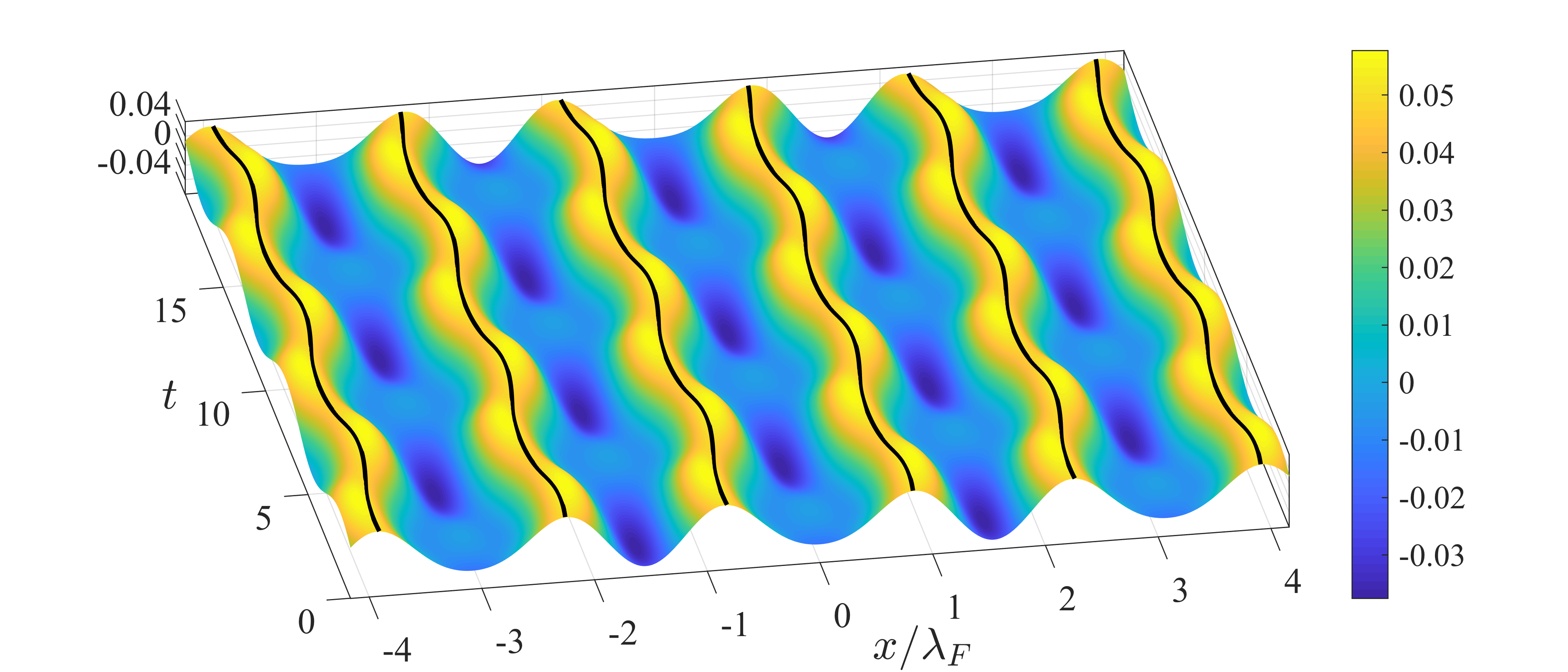}
\end{center}
\caption{Surface plot of the evolution of the wave field, $h$, and droplet positions (black curves) for $N = 22$ droplets, exhibiting out-of-phase oscillations slightly beyond the onset of a supercritical bifurcation (specifically, $\nu_c - \nu = 0.02$).}
\label{fig:wave_time}
\end{figure}

The results of \S\ref{sec:ls} suggest that, for geometrically stable lattice configurations (excluding the case $N = 1$), the lattice destabilises \emph{via} a Hopf bifurcation. (Technically, these are not Hopf bifurcations in the strict sense, as---due to translational invariance of the lattice---the fixed point is non-isolated.) In general, the Hopf bifurcation is one of two types: supercritical, in which stable, small-amplitude oscillations arise beyond the instability threshold, propelled by the slope of the droplet wave field (see Figure \ref{fig:wave_time}); or subcritical, where the system jumps to a distant attractor (such as a solitary-like wave \cite{thomson2020collective}). We proceed to perform a weakly nonlinear analysis in the vicinity of the bifurcation point, $\nu = \nu_{c} = 1/M_c$, allowing us to distinguish between these two contrasting dynamics of this spatially- and temporally-nonlocal system.

When $\nu$ is only slightly less than $\nu_{c}$, we set $\nu_{c} - \nu = \varepsilon^2$, where $0 < \varepsilon\ll 1$, corresponding to $M$ slightly above the critical threshold, $M_{c}$. We then pose the following asymptotic multiple-scales expansions
\begin{equation}
\label{eqn:wnl_expands}
x_{n}(t,T) \sim n\delta + D(T) + \sum_{i=1}^{\infty}\varepsilon^{i} x^{(i)}_{n}(t,T),\quad h(x,t,T) \sim \sum_{i=0}^{\infty}\varepsilon^{i} h^{(i)}(x,t,T),
\end{equation}
where $T = \varepsilon^2 t$ is the slow time-scale. The $O(1)$ drift, $D(T)$, may give rise to a net rotation of the lattice. The origin of this drift may be traced back to the $k = 0$ mode of the dispersion relation \eqref{eqn:disp_rel}, corresponding to translational invariance. (We note that the Stokes' drift in surface gravity waves arises in a similar manner \cite{stokes1880theory}.)

The path-memory stored within the free surface (entering through equation \eqref{eqn:strb_7}) means that the requisite calculations involved in this procedure are non-standard and (unfortunately) lengthy, with full details outlined in Appendix \ref{sec:sl_derive}. Stated succinctly, the outcome of this analysis is that each droplet evolves according to
$$
x_{n}(t,T) = n\delta + \left[D(T) + O(\varepsilon)\right] + \varepsilon\left[A(T)\mye^{\myi(k_{c} n \alpha + \omega_{c}t)} + \text{c.c.}\right] + O(\varepsilon^2),
$$
where the slowly varying complex amplitude $A$ is governed by the Stuart-Landau equation
\begin{equation}
\label{eqn:sl}
\sdone{A}{T} = \gamma_{1} A -\gamma_{2} |A|^2 A.
\end{equation}
The accompanying equation governing the evolution of the drift, $D$, is
\begin{equation}
\label{eqn:D0}
\sdone{D}{T} = \gamma_3|A|^2.
\end{equation}
The coefficients $\gamma_{1},\gamma_{2}\in\mathbb{C}$ and $\gamma_{3}\in \mathbb{R}$ are defined in terms of the system parameters in Appendix \ref{sec:sl_derive}. 

When the critical wavenumber of instability is $k_{c} = N/2$, corresponding to out-of-phase oscillations of the droplets (recall \S\ref{sec:ls}\ref{sec:hopf}), we find that $\gamma_{3} = 0$ and hence $D = \text{constant}$. Otherwise the lattice is endowed with a non-zero drift velocity superimposed on top of individual droplet oscillations, a phenomenon also observed in free-space rings \cite{couchman2020rings}. Further, we observe that when $A$ is constant, $D$ is linear in $T$ and the lattice rotates with constant speed.

Furnished with equation \eqref{eqn:sl}, we now specify to which variety of Hopf bifurcation each lattice configuration destabilises. By recasting $A$ in \eqref{eqn:sl} in polar form $A = \rho\exp(\myi \phi)$ (where $\rho(T)\ge 0$ and $\phi(T)$ are both real) and equating real and imaginary parts, we find
\begin{align}
\label{eqn:rho}
\sdone{\rho}{T} &= r_{1}\rho - r_{2}\rho^3,\\
\label{eqn:phi}
\sdone{\phi}{T} &= s_{1} - s_{2}\rho^2,
\end{align}
where $r_{i} = \text{Re}(\gamma_{i})$ and $s_{i} = \text{Im}(\gamma_{i})$. Consistent with the linear stability analysis performed in \S\ref{sec:ls}, $r_1$ satisfies $r_{1} > 0$, rendering the fixed point $\rho = 0$ (corresponding to a stationary lattice) unstable. If $r_{2}> 0$, there is a second, stable, fixed point $\rho_{*} = \sqrt{r_{1}/r_{2}}$, corresponding to a supercritical Hopf bifurcation. From \eqref{eqn:phi}, the corresponding phase is $\phi(T) = \phi_{*}T$
where $\phi_{*} = s_{1} - s_{2}r_{1}/r_{2}$ (we set the constant of integration, corresponding to an arbitrary phase-shift, to zero by temporal invariance). Conversely, if $r_{2} < 0$ then no additional fixed points emerge beyond the instability threshold, characteristic of a subcritical bifurcation.

We return our attention to Figure \ref{fig:supsub}, in which we mark supercritical ($r_{2} > 0$) and subcritical ($r_{2} < 0$) Hopf bifurcations for varying $N$. Although the precise details of this figure depend on the form of the wave kernel, $\Hkern$, we see that it portrays one of the fundamental features of the experiments \cite{thomson2020collective}, namely the transition from a super- to sub-critical Hopf bifurcation as the number of droplets is doubled from $N = 20$ to $N = 40$.  Physically, stable, small-amplitude oscillations arise for $M > M_c$ when the Hopf bifurcation is supercritical, yet a distant attractor is approached for a subcritical Hopf bifurcation (which, in experiments, takes the form of a solitary wave \cite{thomson2020collective}).  The subtle implications of instead varying the ring radius, $r_{0}$, while keeping the number of droplets, $N$, fixed, are discussed in \S\ref{sec:vary}.

Finally, we justify the efficacy of \eqref{eqn:sl} and \eqref{eqn:D0} in describing the dynamics of the system \eqref{eqn:strb_dimensionless} beyond a supercritical bifurcation point. There is a stable, periodic regime in which each droplet evolves according to
\begin{equation*}
x_{n}(t) = n\delta + d(t) + a\cos(\Omega t + \Psi_{n}) + O(\varepsilon^2),
\end{equation*}
where the drift $d(t) = v(t - t_{0}) + O(\varepsilon)$ proceeds with velocity $v = \gamma_{3}(r_{1}/r_{2})(\nu_c - \nu)$. The amplitude of the oscillations is $a = 2\rho_{*}\sqrt{\nu_{c} - \nu}$ and $\Omega = \phi_{*}(\nu_{c} - \nu) + \omega_{c}$ is the angular frequency. Each oscillation has an associated phase shift $\Psi_{n} = k_{c}n \alpha$. 

\begin{figure}
\begin{center}
\includegraphics[width=\textwidth]{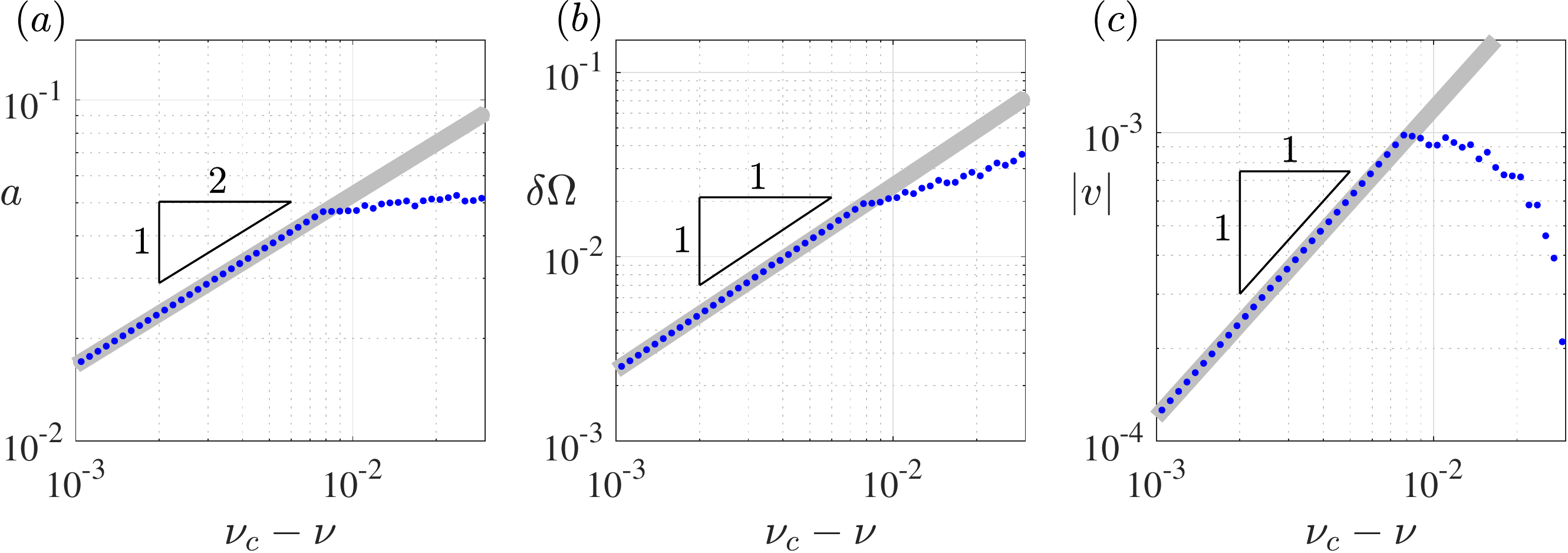}
\end{center}
\caption{Numerical (blue dots) and analytical (grey) predictions for a supercritical Hopf bifurcations for $N =23$, as a function of $\nu_{c} - \nu = \varepsilon^2$. 
$(a)$ The oscillation amplitude $a$. 
$(b)$ The change in angular frequency $\delta\Omega = |\Omega - \omega_c|$. $(c)$ The drift speed $|v|$.
All numerical results were obtained by simulating \eqref{eqn:strb_dimensionless} until a constant-amplitude, periodic state was attained. The numerical and analytical predictions are virtually superimposed when $\varepsilon$ is sufficiently small: as $\varepsilon$ increases, however, the oscillations undergo a second bifurcation, beyond which the dynamics become chaotic in this case.
}
\label{fig:verify}
\end{figure}

In Figure \ref{fig:verify}, we compare direct numerical simulations of \eqref{eqn:strb_dimensionless} with our predictions for $a$, $\Omega$, and $v$ in the supercritical case $N = 23$. Our numerical solutions were computed using a Fourier spectral method, with code and documentation provided in the Supplementary Material. For $0 < \nu_{c} - \nu \ll 1$, our analytical and numerical predictions are virtually superimposed. If $\nu_{c} - \nu$ becomes too large, however, a significant discrepancy arises, which we attribute to the onset of a second bifurcation wherein the oscillations themselves destabilise. It is then necessary to take into account spatial, as well as temporal, variations of the amplitude $A(T)$, which will be explored elsewhere \cite{thomson2020ginzburg}.

\section{Control of stability}
\label{sec:vary}
When the lattice radius, $r_{0}$, is fixed and the number of droplets, $N$, is varied discretely, Figure~\ref{fig:supsub} shows no discernible pattern in the distribution of geometric and memory-driven instabilities. However, the discrete variation of $N$ accounts only for a finite subset of the infinite family of possible droplet separations. To elucidate the underlying structure of the geometric and memory-driven instabilities, we instead consider how the stability of a lattice of $N = 20$ droplets changes as $r_0$ is varied continuously. Herein we characterise the dynamics by the dimensionless separation distance, $\delta = \Lambda/N = 2\pi r_0/N$. As we shall see, regions of geometric and memory-driven instability in fact arise quasi-periodically, with a period close to the Faraday wavelength. (We recall that all lengths are normalised by $\lambda_{F}$.) We note that the wave kernel, $\Hkern(x)$, depends on $r_0$ through equation \eqref{eqn:wavefield_nd}.
\begin{figure}
\begin{center}
\includegraphics[width=\textwidth]{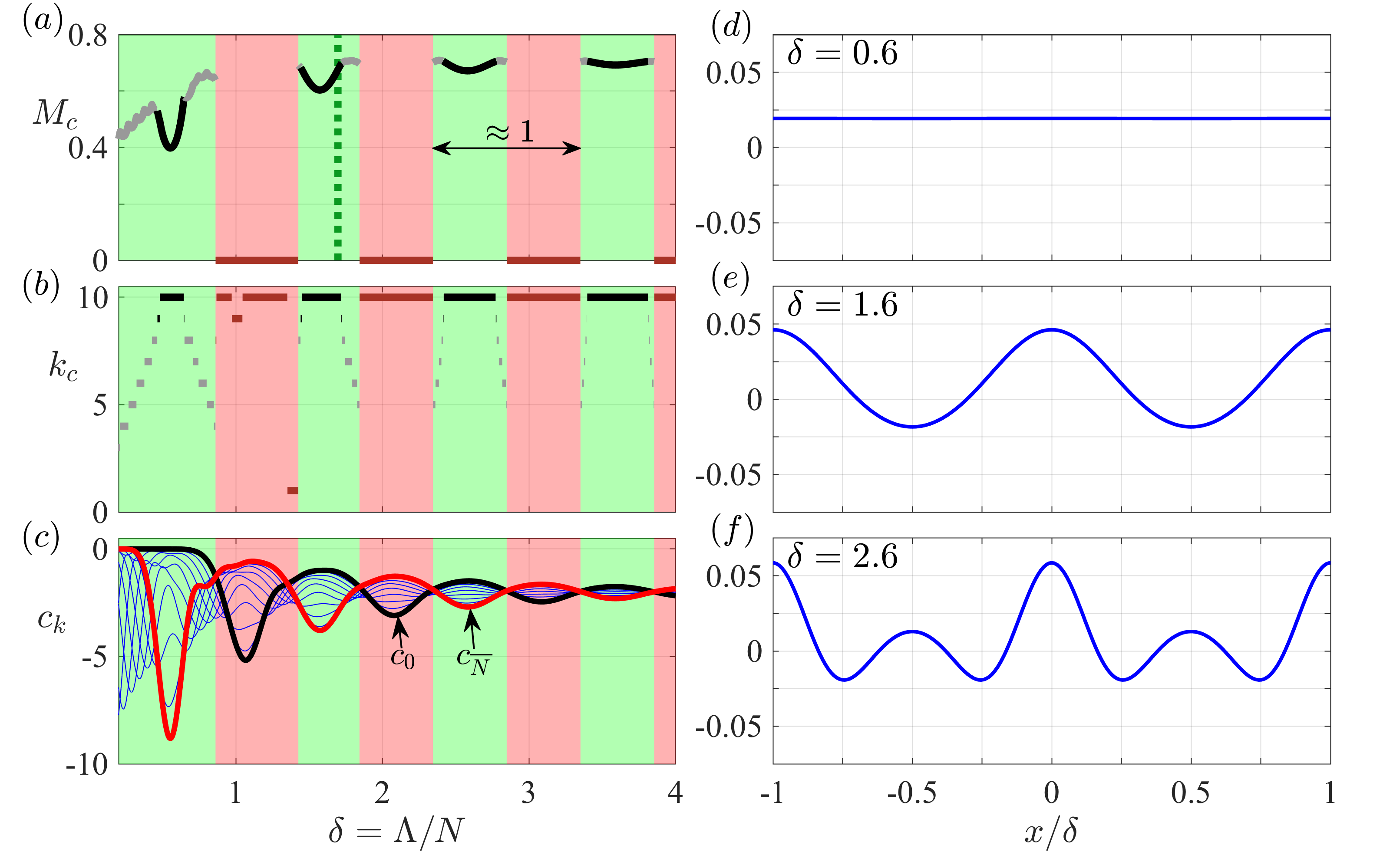}
\end{center}
\caption{Delineation of the instability form when $N = 20$ is fixed and the dimensionless ring radius, $r_0 = \Lambda/(2\pi)$, is varied, altering the dimensionless separation distance, $\delta = \Lambda/N$. In $(a)$--$(c)$, green and red backgrounds correspond to memory-driven $(M_{c} > 0)$ and geometric instabilities $(M_{c} = 0)$, respectively.
$(a)$ The critical memory, $M_c$, where the lattice is unstable for $M > M_c$. Kinks in $M_c$ correspond to jumps in $k_c$. The dashed curve (dark green) corresponds to the $N = 20$ instability in Figure \ref{fig:supsub}.
$(b)$ The corresponding critical wavenumber, $k_c$. For subcritical bifurcations, there is a notable departure in $k_{c}$ from $k = \overline{N}$, an apparent hallmark of such configurations. In both $(a)$ and $(b)$, black and grey curves denote super-  and sub-critical Hopf bifurcations, respectively. 
Dark red lines correspond to geometric instability.
$(c)$ The coefficients $c_k$ (blue curves), where $c_0$ and $c_{\Nbar} = c_{10}$ are highlighted in black and red, respectively.
$(d)$--$(f)$ Steady lattice wave field centred beneath a droplet, extending to the two neighbouring droplets at $\delta = \pm 1$, for 
$(d)$ $\delta = 0.6$,
$(e)$ $\delta = 1.6$, and
$(f)$ $\delta = 2.6$,
all corresponding to supercritical Hopf bifurcations. When the droplets are packed tightly together, oscillations in the steady wave field become imperceptible.
 }
\label{fig:vary_r0}
\end{figure}

\subsection{Geometric instability}

We first study the origins of geometric instability, which we recall from \S\ref{sec:ls}\ref{sec:c3} occurs if there exists an integer $k_*$ such that $c_{k_*} > c_0$. To aid the following discussion, we plot the coefficients $c_{k}$ for $k\in[0,\overline{N} = 10]$ in Figure \ref{fig:vary_r0}(c) as a function of $\delta$. The oscillatory form of the coefficients $c_{k}$ with increasing $\delta$ (or $r_{0}$) arises due to the quasi-monochromatic form of $\Hkern(x)$, the oscillation length-scale being close to the Faraday wavelength (unity in dimensionless units). Moreover, $c_0$ and $c_{\Nbar}$ tend to oscillate out of phase and bound the oscillations of $c_1, \ldots, c_{\Nbar-1}$. The result is that regions of geometric instability typically begin and end at those values of $\delta$ where $c_{\overline{N}}$ crosses $c_{0}$ from below and above, respectively, thus selecting $k_{*} = \overline{N}$. Moreover, each successive region of geometric instability is separated by a Faraday wavelength ($\delta\approx 1$). This effect is preserved even when the droplets are well-separated and interact weakly, specifically when $\delta$ greatly exceeds the wave kernel's decay-length ($\delta \gg l$). In this limit, the coefficients $c_k$ decay exponentially (to the asymptotic value $\mathcal{H}''(0)$, that of a single isolated droplet) over a length scale close to $l$, yet remain oscillatory.
 
\subsection{Memory-driven instability} 
In intervals of $\delta$ where the system undergoes a memory-driven Hopf bifurcation, Figure \ref{fig:vary_r0}(a) demonstrates that the instability threshold, $M_{c}$, varies smoothly, except at points where $k_c$ switches between different integer values (see Figure \ref{fig:vary_r0}(b)). It appears the Hopf bifurcation is supercritical \emph{only} when $k_c$ is $\Nbar$ or $\Nbar - 1$ and subcritical otherwise (recall Figure \ref{fig:fig1}). When the droplets lie less then than a Faraday wavelength apart ($\delta \lesssim 1$), the supercriticality region is very narrow, explaining the prevalence of subcritical bifurcations when the droplets are tightly packed. 
We note for $\delta \lesssim 0.7$ in the current parameter regime (Figure \ref{fig:vary_r0}$(d)$), the steady lattice wave field exhibits imperceptible oscillations, which present a marked contrast to when $\delta \gtrsim 1$ (Figures \ref{fig:vary_r0}(e) and (f)), for which supercritical Hopf and geometric instabilities are prevalent.

\subsection{Summary}

The regions of geometric instability thus provide quantised separation distances prescribed by the Faraday wavelength at which oscillations arise beyond the instability threshold. Moreover, we infer that subcritical bifurcations (and hence solitary-like waves) are more likely to arise when the droplets are either packed closely together ($\delta\lesssim 1$) or when $\delta$ is near the boundaries of geometric instability. However the solitary wave dynamics are not captured by our highly simplified model, likely due to the omission of several potentially significant fluid mechanical effects that arise in this regime (see \S\ref{sec:model}). Nevertheless, the foregoing observations are entirely consistent with experiments \cite{thomson2020collective}, where small-amplitude, out-of-phase oscillations and solitary-like waves are observed when the droplet spacings are around $1.8\lambda_{F}$ and $0.9\lambda_{F}$, respectively. Finally, the foregoing results point to the sensitivity of the experimental system \cite{thomson2020collective}: we infer from Figure \ref{fig:vary_r0} that a small change in ring radius, $r_{0}$, or the addition/subtraction of a droplet may adversely affect the stability of the lattice, for example by pushing the lattice into a region of geometric instability.

\section{Conclusion}
\label{sec:con}
In this paper, we considered the onset and stability of collective vibrations in an active hydrodynamic lattice. This new class of dynamical oscillator is characterised by two features atypical of traditional oscillator systems: spatially nonlocal coupling, mediated in this case by the droplet wave field; and memory-driven self-propulsion of the lattice components. A consequence of the spatially nonlocal coupling is the possibility of geometric instability, where the lattice wave field acts to propel the droplets for all values of the memory parameter, $M$. Otherwise, linear stability predicts the system typically destabilises to an oscillatory instability beyond a critical memory.

The fate of the system close to the point of instability was then studied \emph{via} a systematic weakly nonlinear analysis. At the onset of a memory-driven instability, the lattice system undergoes a Hopf bifurcation, which can be either supercritical, prompting stable, small-amplitude oscillations beyond the instability threshold, or subcritical, where the system jumps to a distant attractor, manifest in experiments as the excitation of a solitary-like wave \cite{thomson2020collective}. In the supercritical regime, an oscillatory-rotary motion may arise when the critical wavenumber $k_{c}\neq N/2$, a phenomenon observed in confined and free-space rings \cite{thomson2020collective,couchman2020rings}. Moreover, when the lattice radius, $r_{0}$, was fixed, our weakly nonlinear analysis corroborated one of the fundamental features of the experiments \cite{thomson2020collective}, namely the transition from a super- to sub-critical Hopf bifurcation as the number of droplets was doubled from 20 to 40.

The dependence of the stability of the lattice on the system parameters was probed yet further when we allowed the ring radius to vary continuously, while keeping the number of droplets fixed. Here we found quantised separations distances similar in size to the Faraday wavelength, separating regions of geometric and memory-driven instability. Generalising our result of the transition from a super- to sub-critical Hopf bifurcation, we infer that subcritical bifurcations are more likely to arise when droplets are more tightly bound (less than a Faraday wavelength apart), while supercritical and geometric instabilities prevale when the droplet spacing is increased. These observations are consistent with the experimental results obtained for the droplet spacings reported in \cite{thomson2020collective}.

The potential avenues for investigation prompted by the results of this paper are numerous, and so we name only a few of the most salient here. Perhaps the most demanding is a description of the fully nonlinear dynamics of the solitary wave regime, specifically extending our model to consider variations in the droplets' vertical bouncing phase \cite{couchman2019bouncing}, or two-dimensional effects imposed by the geometry of the submerged annular channel \cite{durey2020faraday}. Indeed, the fact that the wave force vanishes for tightly packed, equispaced lattices (Figure \ref{fig:vary_r0}(d)) is a portent that the simplified model \eqref{eqn:strb_dimensionless} is insufficient to capture the solitary wave dynamics. Inspired by mathematical models of traffic flow \cite{flynn2009self}, another avenue is to course-grain the discrete model \eqref{eqn:strb_dimensionless} to derive a system of partial differential equations for the macroscopic droplet density and velocity. Such models provide a natural framework in which to study the propagation of nonlinear waves. 

Our system also begs further attention when viewed as a dynamical system. In the future \cite{thomson2020ginzburg}, we will show that taking into account spatial variations in the droplet oscillation amplitude, $A$, genralises equation \eqref{eqn:sl} to a complex Ginzburg-Landau equation, allowimg us to rationalise the onset of the second bifurcation alluded to in Figure \ref{fig:verify}. The techniques employed in this study may also, in principle, be extended to two-dimensional lattices \cite{protiere2005self,lieber2007self,eddi2009archimedean,eddi2011oscillating,couchman2020rings}. Finally, the spatio-temporal nonlocality present in our system also renders it a tantalising experimental \cite{hagerstrom2012experimental,tinsley2012chimera,nkomo2013chimera,totz2018spiral} and theoretical \cite{kuramoto2002coexistence,shima2004rotating,abrams2004chimera, abrams2006chimera,sethia2008clustered} candidate to investigate so-called chimera states in coupled oscillators.

\newpage
\begin{appendices}
\section{Derivation of the Stuart-Landau and drift equations}
\label{sec:sl_derive}
In this section, we provide details of the multiple-scales expansion leading to the amplitude and drift equations \eqref{eqn:sl} and \eqref{eqn:D0}. The basic recipe is thus: first, we substitute the asymptotic expansions \eqref{eqn:wnl_expands} into the stroboscopic model \eqref{eqn:strb_dimensionless} and gather successive powers of $\varepsilon$. At each order, we suppress resonant terms (those that are either constant in $t$ or proportional to $\mye^{\myi\phi_{n}(t)}$, where $\phi_{n} = k_{c}n\alpha + \omega_{c}t$), which are thereby solutions of the linear problem \eqref{eqn:ls6}. This step is facilitated by introducing auxiliary variables to solve for the free surface $h$, akin to \S\ref{sec:ls}. This procedure gives rise to the Stuart-Landau equation \eqref{eqn:sl} at $O(\varepsilon^3)$ and the drift equation \eqref{eqn:D0} at $O(\varepsilon^2)$.

At leading order we obtain a similar system to \eqref{eqn:ls1}:
\begin{equation}
\label{eqn:wn1}
\pd{\hzero}{x}\bigg\vert_{x = \xzero_{n}} = 0,\quad \hzero(x,T) = \frac{1}{\nu_{c}}\sum_{m = 1}^{N}\Hkern(x - \xzero_{m}),
\end{equation}
where $\xzero_{n} = \xzero_{n}(T) = n\delta + D(T)$. By symmetry of $\hzero$, all odd-derivatives vanish beneath each droplet at equilibrium, a fact we will make repeated use of in simplifying forthcoming terms in our expansion.

At $O(\varepsilon)$, we obtain an analogous problem to \eqref{eqn:L1} for $\xone_{n}$ and $\hone$. Hence, after defining the auxiliary variables $X_{n}$ satisfying
$$
\pd{X_{n}}{t}+ \nu_{c}X_{n} = \xone_{n},
$$ 
we obtain the particular solution
\begin{equation}
\label{eqn:wn3}
\hone = -\sum_{m=1}^{N}X_{m}\Hkern'(x - \xzero_{m})
\end{equation}
and $\mathcal{L}_{n}\bm{x}^{(1)} = 0$, where the definition of the linear operator, $\mathcal{L}_{n}$, is the same as in \eqref{eqn:ls6} (with $\nu = \nu_{c}$) and $\bm{x}^{(1)} = (x^{(1)}_{1},\ldots,x^{(1)}_{N})$. (Recall that the homogeneous part of $h^{(1)}$ decays exponentially in time (\S\ref{sec:ls}).) We now seek a solution to $\mathcal{L}_{n}\bm{x}^{(1)}= 0$ of the form
\begin{align}
\xone_{n} = B(T) + \left[A(T)\mye^{\myi\phi_{n}} + \text{c.c.}\right],\quad
X_{n} = \frac{1}{\nu_{c}}B(T) + \left[\frac{A(T)}{\nu_{c} + \myi\omega_{c}}\mye^{\myi\phi_{n}} + \text{c.c.}\right],
\end{align}
where $A$ is a complex amplitude and $B$ is a correction to the drift, $D$. This solution may be verified by recalling that the dispersion relation, $\mathcal{D}_{k}$, satisfies $\mathcal{D}_{k_{c}}(\myi\omega_{c}; \nu_{c}) = 0$ and $\mathcal{D}_{0}(0; \nu) = 0$.

At $O(\varepsilon^2)$, we have the following system for $\xtwo_{n}$ and $\htwo$:
\begin{align}
\label{eqn:xn2}
\pdd{2}{\xtwo_n}{t} + \pd{\xtwo_n}{t} + \sdone{D}{T} &= -\left\{\pd{\htwo}{x} + \xtwo_{n}\frac{\partial^2 h^{(0)}}{\partial x^2} + \xone_{n}\frac{\partial^2\hone}{\partial x^2}\right\}\Big\vert_{x = \xzero_{n}},\\
\label{eqn:h2}
\frac{\partial\htwo}{\partial t} + \nu_{c}\htwo + \pd{\hzero}{T} &= \hzero+
\sum_{m=1}^{N}\left\{\frac{1}{2}x^{(1)2}_{m}\Hkern''(x - \xzero_{m})-\xtwo_{m} \Hkern'(x - \xzero_{m})\right\}.
\end{align}
To solve for $\htwo$, we introduce two further auxiliary variables, $Y_{n}$ and $Z_{n}$, akin to the procedure adopted in \eqref{eqn:ls4}. By the form of the inhomogeneity in \eqref{eqn:h2}, we pose that $Y_{n}$ and $Z_{n}$ satisfy
\begin{equation}
\label{eqn:X2_X3}
\pd{Y_{n}}{t} + \nu_{c}Y_{n} = \frac{1}{2}x^{(1)2}_{n} ,\quad \pd{Z_{n}}{t} + \nu_{c}Z_{n} = \xtwo_{n}.
\end{equation}
A particular solution of \eqref{eqn:h2} is then found:
\begin{equation}
\label{eqn:h2_sol}
\htwo = \frac{1}{\nu_{c}}\left(\hzero - \pd{\hzero}{T}\right) +\sum_{m=1}^{N}\left\{Y_{m}\Hkern''(x - \xzero_{m}) - Z_{m}\Hkern'(x - \xzero_{m})\right\}.
\end{equation}
Using the form of $\xone_{n}$, we find from the first equation in \eqref{eqn:X2_X3} that
\begin{equation}
\label{eqn:Xtwo}
Y_{n} = \frac{1}{\nu_{c}}\left[|A|^2 + \frac{1}{2}B^2\right] + B\left[\frac{A}{\nu_{c} + \myi\omega_{c}}\mye^{\myi\phi_{n}}+\text{c.c.}\right] + \frac{1}{2}\left[\frac{A^2}{\nu_{c} + 2\myi\omega_{c}}\mye^{2\myi\phi_{n}} + \text{c.c.}\right].
\end{equation}
Substituting \eqref{eqn:h2_sol} and \eqref{eqn:Xtwo} into \eqref{eqn:xn2}, and using \eqref{eqn:wn1}, then yields
\begin{equation}
\label{eqn:L_x2}
\mathcal{L}_{n}\bm{x}^{(2)} = -\alpha_{0}\sdone{D}{T} + \left\{a_{1}A^2\mye^{2\myi\phi_{n}} + \text{c.c.}\right\} + a_{2}|A|^2,
\end{equation}
where the coefficients $\alpha_{0}$, $a_{1}$, and $a_{2}$ are included in the summary at the end of this section. 

For a bounded solution of \eqref{eqn:L_x2}, we require that the constant secular terms on the right-hand side vanish, yielding an equation governing the drift $D$:
\begin{equation}
\label{eqn:driftD}
\alpha_{0}\sdone{D}{T} = a_{2}|A|^2.
\end{equation}
As terms proportional to $\mye^{\pm 2\myi\phi_{n}(t)}$ in \eqref{eqn:L_x2} are non-secular, the general solution of \eqref{eqn:L_x2} is  then
\begin{equation*}
\xtwo_{n} = \left[C(T)\mye^{\myi\phi_{n}} + \text{c.c.}\right] + \left[\frac{a_{1}A^2}{\mathcal{D}_{2k_{c}}(2\myi\omega_{c};\nu_{c})}\mye^{2\myi\phi_{n}} + \text{c.c.}\right] + E(T),
\end{equation*}
where $C$ and $E$ are corrections to complex amplitude, $A$, and drift, $D$, respectively. 
Thus, from the second equation in \eqref{eqn:X2_X3}, we find
\begin{equation*}
Z_{n} = \left[\frac{\tilde{a}_{1}A^2}{\nu_{c} + 2\myi\omega_{c}}\mye^{2\myi\phi_{n}} + \text{c.c.}\right] + \left[\frac{C}{\nu_{c} + \myi\omega_{c}}\mye^{\myi\phi_{n}} + \text{c.c.}\right] + \frac{1}{\nu_{c}}E,\quad\text{where}\quad \tilde{a}_{1} = \frac{a_{1}}{\mathcal{D}_{2k_{c}}(2\myi\omega_{c};k_{c})}.
\end{equation*}

At $O(\varepsilon^3)$, we have a system for $\xthree_{n}$ and $\hthree$, namely
\begin{multline}
\label{eqn:xthree}
\pdd{2}{\xthree_{n}}{t} + \pd{\xthree_{n}}{t} + \xthree_{n}\frac{\partial^2 \hzero}{\partial x^2}\bigg\vert_{x = \xzero_{n}} = -\left[2\frac{\partial^2\xone_{n}}{\partial t\partial T} + \pd{\xone_{n}}{T}\right]\\ - \left[\frac{\partial h^{(3)}}{\partial x} + \xone_{n}\frac{\partial^2 \htwo}{\partial x^2} + \frac{1}{2}x^{(1)2}_{n}\frac{\partial^3 \hone}{\partial x^3} + \xtwo_{n}\frac{\partial^2 \hone}{\partial x^2} + \frac{1}{6}x^{(1)3}_{n}\frac{\partial^4\hzero}{\partial x^4}\right]\Big\vert_{x = \xzero_{n}},
\end{multline}
\begin{multline}
\label{eqn:hthree}
\pd{\hthree}{t} + \nu_{c}\hthree = -\left[\pd{\hone}{T} - \hone\right] \\+ \sum_{m=1}^{N}\bigg\{\xone_{m}\xtwo_{m}\Hkern''(x - \xzero_{m}) - \xthree_{m}\Hkern'(x - \xzero_{m}) - \frac{1}{6}x^{(1)3}_{m}\Hkern'''(x - \xzero_{m})\bigg\}.
\end{multline}
In an identical procedure to that carried out at $O(\varepsilon^2)$, introducing three further auxiliary variables allows us to find a particular solution of \eqref{eqn:hthree} governing $\hthree$. Then, after substituting $\hthree$ into \eqref{eqn:xthree}, eliminating secular terms (either constant in $t$ or proportional to $\mye^{\myi\phi_{n}(t)}$) yields two equations governing the complex amplitude, $A$, and real drift, $B$, namely
\begin{align}
\label{eqn:sll}
\alpha _1\sdone{A}{T} &= \alpha_2 A -\alpha_3 |A|^2 A, \\
\label{eqn:oedrift}
\alpha_0\sdone{B}{T} &= 2a_2\,\mathrm{Re}[A^*C],
\end{align}
in addition to equation \eqref{eqn:driftD} governing the drift, $D$. The notation $A^{*}$ denotes the complex conjugate of $A$.
We cannot determine the higher-order corrections $B$, $C$, and $E$ without proceeding to $O(\varepsilon^4)$ and higher. Nevertheless, a satisfactory approximation is obtained by considering $A$ and $D$ alone, as made clear by Figure \ref{fig:verify}.

We conclude by summarising the coefficients, where it helps to recall that $\mathcal{D}_{k}(\lambda;\nu)$ is the dispersion relation. For convenience, we introduce the notation $\Hkern_{n} = \Hkern(n\delta)$ (and similarly for derivatives). The coefficients $\alpha_{0}$, $a_{1}$, and $a_{2}$ appearing in \eqref{eqn:L_x2}, \eqref{eqn:driftD}, and \eqref{eqn:oedrift} are
\begin{align*}
\alpha_0 &= \pd{\mathcal{D}_0}{\lambda}(0;\nu_c) =   1 + \frac{1}{\nu_c^2}\sum_{n = 1}^N \Hkern''_n,\\
a_{1} &= \frac{
1}{\nu_{c} + \myi\omega_{c}}\sum_{n=1}^{N}\mye^{-\myi k_{c}n\alpha}\Hkern'''_{n} - \frac{1}{2(\nu_{c} + 2\myi\omega_{c})}\sum_{n=1}^{N}\mye^{-2\myi k_{c}n\alpha}\Hkern'''_{n},\\
a_{2} &= 2\text{Re}\left[\frac{1}{\nu_{c} + \myi\omega_{c}}\sum_{n=1}^{N}\mye^{-\myi k_{c}n\alpha}\Hkern'''_{n}\right] = -\frac{2\omega_{c}}{\nu^2_{c} + \omega^2_{c}}\sum_{n=1}^{N}\sin(k_{c}n\alpha)\Hkern'''_{n},
\end{align*}
while the $\alpha_{i}$, for $i = 1,2,3$, appearing in \eqref{eqn:sll} are
$$
\alpha_1 = \pd{\mathcal{D}_{k_c}}{\lambda}(\myi\omega_c; \nu_c),\quad 
\alpha_2 = \pd{\mathcal{D}_{k_c}}{\nu}(\myi\omega_c; \nu_c),
$$
and 
\begin{multline*}
\alpha_3 = \frac{3}{2\nu_c}\sum_{n=1}^N \Hkern''''_n - \sum_{n=1}^N\Hkern''''_n \mathrm{Re}\bigg[\frac{\mye^{-\myi k_c n \alpha}}{\nu_c + \myi \omega_c}\bigg] + 
\frac{1}{2}\sum_{n=1}^N \frac{\Hkern''''_n\mye^{-2\myi k_c n \alpha}}{\nu_c + 2\myi\omega_c}-
\sum_{n=1}^N \frac{\Hkern''''_n\mye^{-\myi k_c n \alpha}}{\nu_c + \myi\omega_c} \\
+{2\myi \tilde{a}_{1}\sum_{n=1}^N\Hkern'''_n \mathrm{Im}\bigg[\frac{\mye^{-\myi k_c n \alpha}}{\nu_c + \myi \omega_c}\bigg] - \tilde{a}_{1}\sum_{n=1}^N\frac{\Hkern'''_n\mye^{-2\myi k_c n \alpha} }{\nu_c + 2\myi\omega_c}
- \frac{a_2}{\alpha_0} \sum_{n = 1}^N\frac{\mye^{-\myi k_c n \alpha} \Hkern_n'''}{(\nu_c + \myi \omega_c)^2}}.
\end{multline*}
Recall that $\tilde{a}_{1} = a_{1}/\mathcal{D}_{2k_{c}}(2\myi\omega_{c};\nu_{c})$. On dividing \eqref{eqn:sll} by $\alpha_{1}$ and \eqref{eqn:driftD} by $\alpha_{0}$ we arrive at equations \eqref{eqn:sl} and \eqref{eqn:D0} in the main text, where $\gamma_{1} = \alpha_{2}/\alpha_{1}$, $\gamma_{2} = \alpha_{3}/\alpha_{1}$, and $\gamma_{3} = a_{2}/\alpha_{0}$.
We note that when $k_{c} = N/2$, simplifications arise since $\tilde{a}_{1} = a_{2} = 0$.

\end{appendices}

\bibliography{paperfinal}
\bibliographystyle{unsrt}

\end{document}